\pdfoutput=1
\documentclass[12pt]{iopart}

\usepackage{graphicx}
\usepackage{cite}
\usepackage{url}
\usepackage{booktabs}

\usepackage{lmodern}
\usepackage{gensymb}
\usepackage{textcomp}

\usepackage[letterpaper]{geometry}
\begin{document}

\title{Comb-calibrated solar spectroscopy through a multiplexed single-mode
fiber channel}

\author{R~A~Probst$^{1}$, L~Wang$^{2,\,3}$, H-P~Doerr$^{4}$, T~Steinmetz$^{5}$,
T~J~Kentischer$^{4}$, G~Zhao$^{2}$, T~W~H\"ansch$^{1}$, Th~Udem$^{1}$,
R~Holzwarth$^{1,\,5}$ and W~Schmidt$^{4}$}

\address{$^{1}$Max-Planck-Institut f\"ur Quantenoptik, Hans-Kopfermann-Str.
1, 85741 Garching, Germany\\
$^{2}$Key Laboratory of Optical Astronomy, National Astronomical
Observatories, Chinese Academy of Sciences, A20 Datun Road, Beijing
100012, China\\
$^{3}$Universit\"ats-Sternwarte M\"unchen, Scheinerstr. 1, 81679 M\"unchen,
Germany\\
$^{4}$Kiepenheuer-Institut f\"ur Sonnenphysik, Sch\"oneckstr. 6, 79104
Freiburg, Germany\\
$^{5}$Menlo Systems GmbH, Am Klopferspitz 19a, 82152 Martinsried,
Germany\\
\vspace{1mm}
E-mail: \url{rafael.probst@mpq.mpg.de}}

\begin{abstract}
\noindent We investigate a new scheme for astronomical spectrograph
calibration using the laser frequency comb at the Solar Vacuum Tower
Telescope on Tenerife. Our concept is based upon a single-mode fiber
channel, that simultaneously feeds the spectrograph with comb light
and sunlight. This yields nearly perfect spatial mode matching between
the two sources. In combination with the absolute calibration provided
by the frequency comb, this method enables extremely robust and accurate
spectroscopic measurements. The performance of this scheme is compared
to a sequence of alternating comb and sunlight, and to absorption
lines from Earth's atmosphere. We also show how the method can be
used for radial-velocity detection by measuring the well-explored
5-minute oscillations averaged over the full solar disk. Our method
is currently restricted to solar spectroscopy, but with further evolving
fiber-injection techniques it could become an option even for faint
astronomical targets.

\noindent ~

\end{abstract}

\section{Introduction}

Laser frequency combs (LFCs) have revolutionized precision spectroscopy
in atomic and molecular physics~\cite{Udem02}, enabling measurements
of transition frequencies with unprecedented accuracy. Recently, they
have grown into high-precision frequency calibrators for astronomical
spectrographs~\cite{Murphy07,Steinmetz08,Wilken10,Ycas:12,Phillips:12,Doerr12,Doerr12SPIE,Wilken12,Molaro13}.
An LFC represents an absolute optical frequency reference, directly
linked to an accurate radio-frequency (RF) standard such as an atomic
clock. Its spectrum consists of a very regular series of sharp lines
(or modes) at frequencies $f_{n}=f_{0}+n\, f_{r}$, where the repetition
frequency $f_{r}$ is the mode spacing, the offset frequency $f_{0}$
is a global shift of the mode structure, and the mode number $n$
is an integer. $f_{r}$ and $f_{0}$ are both radio frequencies, accessible
to conventional electronics, while the values of $n$ are high enough
(from about 80\,000 to 120\,000 for the LFC described in this work)
to enable the transfer from RF to optical frequencies. With its unique
properties, the LFC comes close to an ideal calibrator for spectrographs~\cite{Murphy07},
whose outstanding accuracy is expected to open up new perspectives
in astronomy. This comprises the detection of Earth-like extrasolar
planets by measuring the recoil motion of their host stars~\cite{Wilken12}.
LFCs may also enable a direct measurement of the acceleration of the
cosmic expansion~\cite{Liske08}, and a more accurate cosmological
search for variability of fundamental constants~\cite{Webb11}.

A central aspect in precision astronomical spectroscopy is the measurement
of radial-velocity (RV) changes of celestial bodies via Doppler shifts
of spectral lines. With conventional calibration techniques, the best
spectrographs are currently limited to an RV precision of roughly
1~m/s over large time spans. LFCs hold promise to enable RV measurements
at an accuracy of 1~cm/s over arbitrary time horizons, and a 2.5~cm/s
calibration repeatability has already been demonstrated on a time
scale of hours~\cite{Wilken12}. Unlike other calibration sources,
LFCs are not only repeatable, but to the same degree also accurate,
i.e. they allow assigning wavelengths, or optical frequencies, that
come close to the true values. Most applications would profit from
LFCs mainly for their high repeatability, but the high accuracy of
LFCs can also be beneficial, e.g. to compare spectra from different
instruments. This property has also been exploited by~\cite{Molaro13}
to create an improved atlas of the solar lines in the visible.

Current high-precision spectrographs are usually connected to their
telescopes via optical fibers, to detach them from telescope guiding,
increasing the stability of the measurements. The spectrographs are
usually equipped with a second fiber channel that is used to simultaneously
calibrate the spectrograph during each measurement, to track spectrograph
drifts. Differential drifts between the two channels are minimized
by a rigid mechanical connection between the fiber outputs. However,
since it is challenging to couple light from astronomical objects
efficiently into single-mode fibers (SMFs), current high-precision
spectrographs are fed with multimode fibers. While an SMF guides only
a single, approximately Gaussian shaped beam profile, a multimode
fiber supports a variety of beam profiles. The profile at the output
of a multimode fiber depends on the beam alignment at its input, and
therefore on telescope guiding. This translates into uncertainties
in RV measurements.

In this work, the use of SMFs for LFC-based calibration of astronomical
spectrographs is explored. The tests presented here have been made
at the Vacuum Tower Telescope (VTT), a solar telescope in Tenerife,
Canary Islands, that is operated by the KIS. The very high optical
powers available from the Sun allow the use of SMFs to feed the spectrograph
both with sunlight and comb light, tolerating low coupling efficiencies
on the part of the Sun. However, advances in adaptive optics are anticipated
to enable efficient SMF coupling of distant stars~\cite{Jovanovic14},
making our approach interesting for night-time astronomy as well.
As an alternative to adaptive optics, photonic lanterns have recently
been developed as efficient multimode to single-mode converters for
astronomical applications~\cite{Betters:13,Spaleniak:13}. They split
up the light of a multimode fiber into an array of SMF channels. The
use of SMFs fully decouples the spatial beam parameters in the spectrograph
from telescope guiding, which potentially limits the precision of
other fiber-fed spectrographs. SMFs also eliminate the degradation
of the beam profile through formation of laser speckles upon fiber
coupling of coherent light. For multimode fibers, this occurs due
to modal interference. This point is particularly critical for the
LFC, owing to its high degree of coherence.

Another concept that we introduce for spectrograph calibration with
an LFC is multiplexing of the fiber channel, which involves coupling
both comb and sunlight into one single fiber. We investigate both
temporally separated and simultaneous fiber injection of the two sources.
Especially the latter concept is shown to enable very accurate tracking
of spectrograph drifts during measurements. This eliminates the need
for an extra fiber for calibration, that can potentially drift versus
the other one and thereby cause calibration errors. In combination
with the use of SMFs, we obtain nearly perfect spatial mode-matching
between comb and sunlight, and thus an extremely robust calibration.
This is similar to the mode in which iodine absorption cells are operated,
which however absorb a part of the science light, and are surpassed
by LFCs in a number of calibration characteristics such as accuracy,
repeatability, and spectral coverage.

We demonstrate our methodology by observing the 5-minute oscillations
in the integrated solar spectrum, i.e. in the spectrum averaged over
the full solar disk. These oscillations are well understood~\cite{Kosovichev10}
and have been extensively characterized~\cite{Claverie79,Davies14,Kjeldsen08}.
They hence represent a suitable test-bed for our method. The observation
of integrated sunlight, considering the "Sun-as-a-star", is of
great interest for night-time astronomy, as it allows investigating
effects like stellar activity on RV detection. These studies could
not be carried out in such detail on other stars, but are needed for
a more accurate search for extrasolar planets. Of course, our method
is not limited to integrated sunlight, and the VTT telescope allows
placing the SMF tip in the image of the Sun. This results in spatially
resolved, comb-calibrated measurements, supported by the adaptive
optics of the VTT. Currently, this approach is being used to study the center-to-limb
variation of the solar convective blue shift, and is proposed for
characterization of the solar meridional motion~\cite{Doerr12}.

\section{Instruments and observations\label{sec:Instruments-and-observations}}

The observations described and analyzed in this work were made at
the VTT in Tenerife~\cite{Schroeter85}, where an infrared LFC had
already been tested in 2008~\cite{Steinmetz08}. For our tests, we
fed the VTT echelle spectrograph with integrated sunlight from the
auxiliary full-disk telescope ChroTel~\cite{Bethge11,Kentischer08},
located in the same building as the VTT telescope. Within the beam
path of the ChroTel, we installed a lens that produced an about 4~mm
wide image of the pupil of the telescope, in which we placed the tip
of an SMF with a 2.5~\micro{}m core diameter. The lens roughly matched the
numerical aperture (NA) of the sunlight to the fiber NA of 0.13. This
resulted in 800~nW of visible sunlight coupled into the SMF, which
corresponds to a coupling efficiency of 0.1~ppm with respect to the
total solar radiation power collected by the telescope. The fairly
low coupling efficiency is mainly due to the fact, that to an earthbound
observer the Sun $-$ unlike other stars $-$ appears as an extended source
with poor spatial coherence properties. Its light thus consists of
a multitude of spatial modes without a well-defined relative phase.
Only one of these modes can be fully matched, and thereby efficiently
coupled, to the single spatial mode of an SMF at a time. The sunlight
collected by the ChroTel consists of several million spatial modes,
as can be deduced from its optical resolution. This necessarily entails
a poor spatial mode overlap of the guided mode of the SMF with the
incident light field at the pupil image. The problem could be circumvented
by coupling each spatial mode to a different SMF, e.g. with a photonic
lantern, but in our case this is impractical because of the great
number of spatial modes. This is different for distant stars, where
photonic lanterns should represent a viable solution. Further, a diffraction
limited image of a distant star, e.g. obtained by employing adaptive
optics, could enable coupling into an SMF with efficiencies of $>$70~\%~\cite{Jovanovic14}.

The SMF in this setup collects light from every point of the solar
disk, but not with equal efficiency, owing to the approximately Gaussian-shaped
acceptance profile of the SMF. This is because in the pupil image,
light from different locations on the Sun is incident from different
angles. As a result, the observed solar spectrum represents a weighted
average over the solar disk. The rotation of the Sun imprints a $\pm$2~km/s
RV variation across the solar disk. Inaccurate telescope guiding can
consequently distort the RV measurement, by shifting the centroid
of the weighted spatial average towards the blue or red-shifted region.
The specified guiding accuracy of 0.5~arcsec of the ChroTel should
keep the resulting RV errors below approximately 1~m/s. This guiding
accuracy can however only be guaranteed under sufficiently good observation
conditions. It is also important to note, that this sort of RV error
from telescope guiding would not exist, if the technique was used
in stellar night-time astronomy, since distant stars are normally
not spatially resolved.

\begin{figure}
\includegraphics[height=7.5cm]{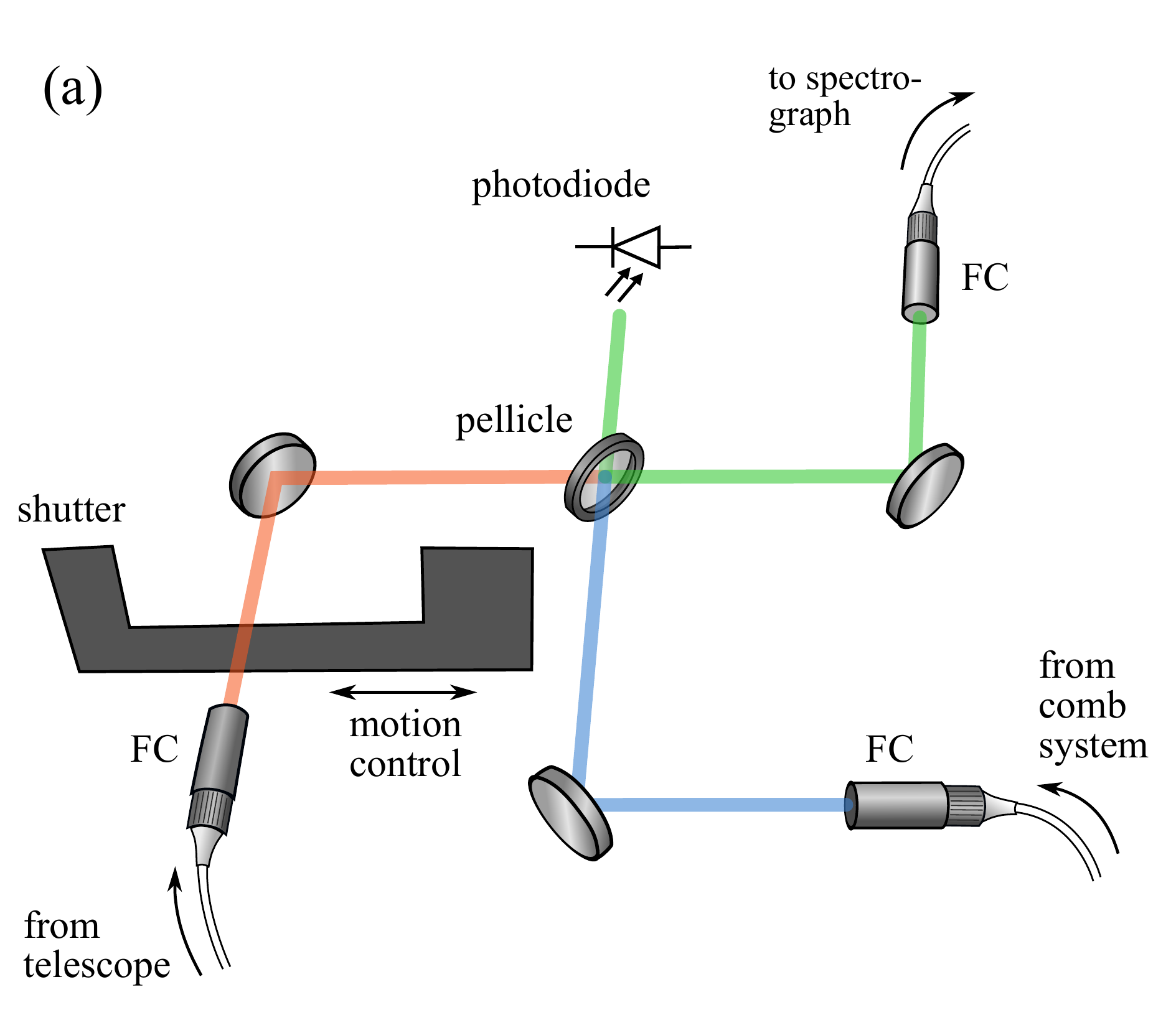}\includegraphics[height=7.5cm]{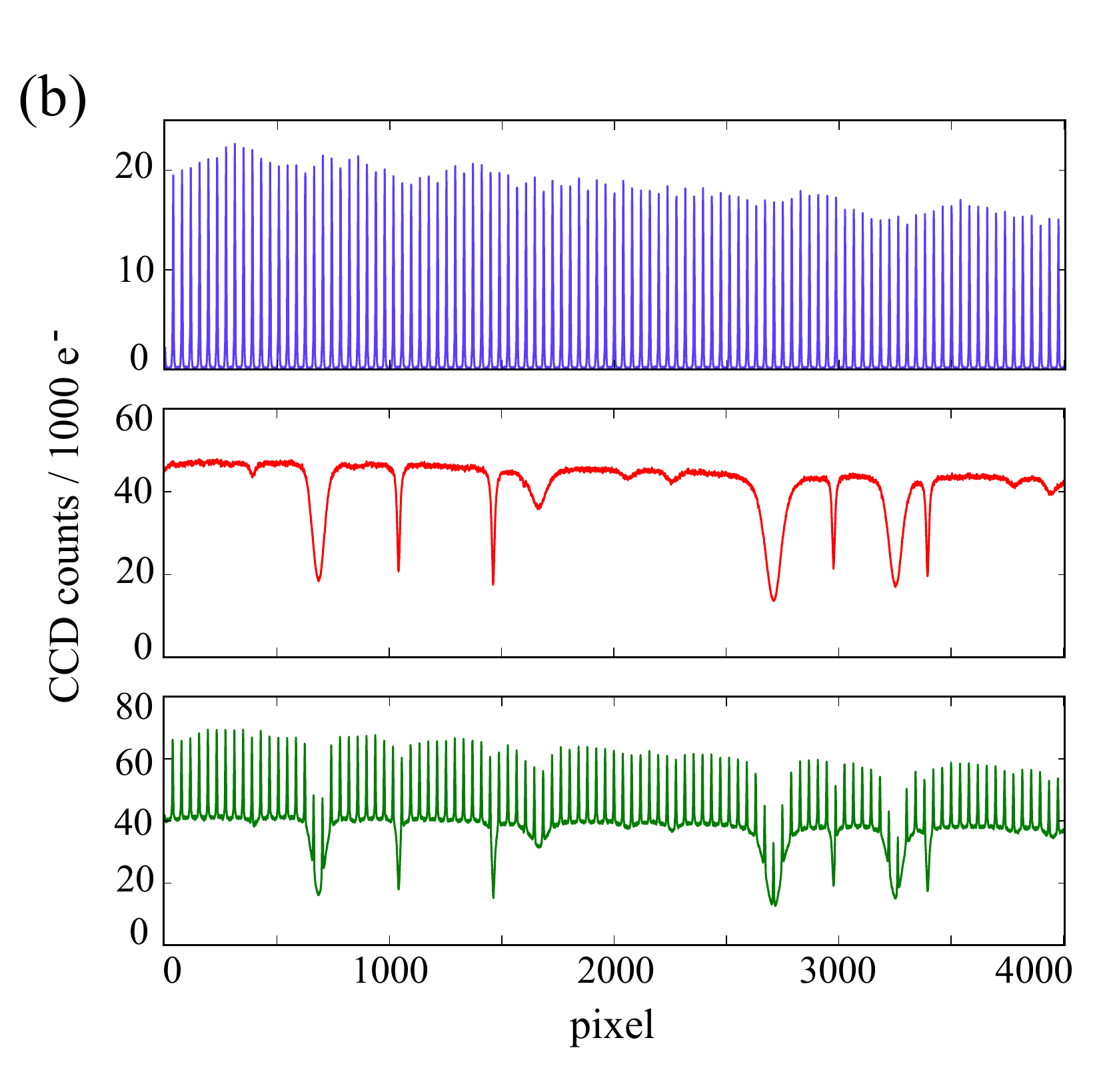}

{\footnotesize{\caption{\label{fig:multiplexer}(a)~Setup of the optical multiplexer: The
sunlight (red) from the telescope and the comb light (blue) enter
the free-space section through fiber collimators (FC). They are combined
(green) on a pellicle beam splitter, and then coupled into the output
fiber leading to the spectrograph. The shutter can be positioned to
block either the sunlight, the comb light, neither, or both. Three
silver mirrors inside the beam path support optical alignment. (b)~Spectra
as measured using the VTT spectrograph at different positions of the
shutter, with the multiplexer transmitting only comb light (top),
sunlight (center), or both (bottom).}
}}
\end{figure}

The LFC as installed at the VTT in October 2011 is described in~\cite{Doerr12}.
In May 2012, just before making the tests reported here, the system
was upgraded to enable higher calibration accuracy and more reliable
operation~\cite{Doerr12SPIE}. The system is referenced to a GPS
disciplined quartz oscillator, accurate within a fractional uncertainty
of $<10^{-12}$ for integration times exceeding 1~s. It features
a broadband visible output coupled into an SMF, and routinely delivers
sufficient power for spectrograph calibration over a 460$-$680~nm spectral
range. It was operated at a repetition frequency of $f_{r}=$~5.445~GHz
and an offset frequency of $f_{0}=$~+100~MHz. This comb spectrum
is synthesized from an LFC with initially smaller mode spacing, by
suppressing unwanted modes with a series of Fabry-P\'erot filters~\cite{Steinmetz09}.
It is estimated, that the unwanted modes are suppressed to better
than 47~dB, and thus cannot shift the calibration by more than 5
kHz~\cite{Probst:13}. At a center wavelength of 630~nm, as used
in this work, this corresponds to 3~mm/s. Furthermore, owing to the
finite width of the comb modes, the mode filtering can shift their
center of gravity by skewing their shape, in case the filter transmission
peaks are not perfectly centered at their position~\cite{Braje08}.
In analogy to the discussion in~\cite{Probst14_SPIE} we estimate,
that this effect can cause line shifts of up to 2.1~cm/s for this
LFC.

The sunlight from the telescope and the calibration light from the
LFC are combined into a one SMF by an optical multiplexer that is
shown in figure~\ref{fig:multiplexer}(a). The fiber-coupled comb
and sunlight enter the free-space section of the multiplexer through
fiber collimators using achromatic triplet lenses. The two beams are
overlapped on a 2~\micro{}m thick uncoated pellicle, that on average reflects
8~\% of the comb light and transmits 92~\% of the sunlight. The
reflectivity (and transmittance) is modulated by interferences
on the two surfaces of the pellicle. Its very low thickness stretches
the period of the modulation to much larger than the observed spectral
range of 0.7~nm (see below). Another solution would be to replace
the pellicle by a wedged glass plate, to suppress the modulation.
The combined beam of comb and sunlight is coupled into an SMF leading
to the spectrograph. All fibers are identical, feature single-mode
guidance from 450 to 680~nm, and use connectors with angled end facets.
The pellicle produces a second overlapped output, which for alignment
purposes is monitored on a photodiode. Within the free-space section,
there is a beam shutter in form of a U-shaped black metal panel, whose
position is controlled with a motorized linear translation stage.
Four positions for the shutter are defined, in which it blocks either
the beam from the Sun, that of the comb, neither, or both.

The output fiber of the multiplexer leads to the VTT echelle spectrograph.
From there, the light is focused through the entrance slit of the
spectrograph by an aspheric lens, that matches the NA of the beam
to that of the spectrograph. The VTT spectrograph has a very high
resolution of $R=10^{6}$, but usually only observes a single echelle
order at a time, and hence only a rather narrow spectral range. The
center wavelength of the observation can be tuned over a wide range
by adjusting the angles of the echelle grating and predisperser. The
image plane of the spectrograph can be accessed freely, allowing to
place and exchange CCD cameras as desired by the user. With the PCO~4000
camera that we used to record our spectra, we could observe a 0.7~nm
wide spectral range that we centered at 630.0~nm. The CCD chip of
the camera was cooled down to $-$18 \celsius{}, had a root-mean-square (RMS)
readout-noise of 11 photoelectrons, and counted 3.3~photoelectrons
for every step of the 14~bit readout amplifier. The CCD had a 9~\micro{}m ~$\times$~9~\micro{}m
pixel pitch with 4008 pixels in the dispersion direction of the echelle
order and 2672 pixels perpendicular to it. We employed an 8-to-1 hardware
binning of the pixels in the latter direction. Along the illuminated
pixels, we summed up 5 of the binned pixels in every column perpendicular
to the dispersion direction, in order to gain a 1-dimensional data
string from the CCD image.

For all exposures, we chose an exposure time of 1000~ms. The position
of the shutter within the multiplexer was changed periodically for
the observations, to first record a spectrum with comb light only,
then with sunlight only, and finally with both, after which the cycle
was repeated (see figure~\ref{fig:multiplexer}(b)). Between the
exposures, we left a 1000~ms pause for the shutter to move. With
this choice of the exposure cycle, two different multiplexed calibration
schemes can be put into practice: If only the frames with pure sunlight
or pure comb light are considered, the solar spectrum can be calibrated
by interpolating the calibrations obtained from the comb exposures
directly before and directly after each Sun exposure. This temporally
separated signal transmission is known as time-division multiplexing.
We refer to this calibration scheme as time-interlaced calibration.
If also the exposures containing both comb and sunlight are considered,
the solar spectra can be calibrated from the overlaid comb. This concept
is hereafter referred to as overlaid calibration. In the following
sections we investigate the benefits and drawbacks of the two calibration
schemes, and assess their performance.

The observations used for our analysis were made on May 29, 2012,
from 12:18 to 17:16 coordinated universal time (UTC). During this
time span, 2715 full cycles as described above were recorded. Unfortunately,
the series was interrupted several times due to technical problems,
and the telescope guiding was partly disturbed by passing clouds.
Towards the end of the series, 86~minutes of uninterrupted operation
with stable guiding were accomplished, which we use as our observed
sample for the comb-Sun comparisons. However, to test the stability
of the comb calibration on its own, all recorded comb spectra are
usable. The observed solar spectrum also contained several $\mathrm{O_{2}}$
lines of telluric origin, i.e. stemming from Earth's atmosphere. Their
line centers are independent of telescope guiding, and hence to compare
the LFC with these lines, all recorded solar spectra~are~of~use.

\section{Calibration tests\label{sec:Comb-calibration}}

\subsection{Spectrograph calibration\label{sub:Spectrograph-calibration}}

\begin{figure}
\begin{indented}\item[]\includegraphics[width=9cm]{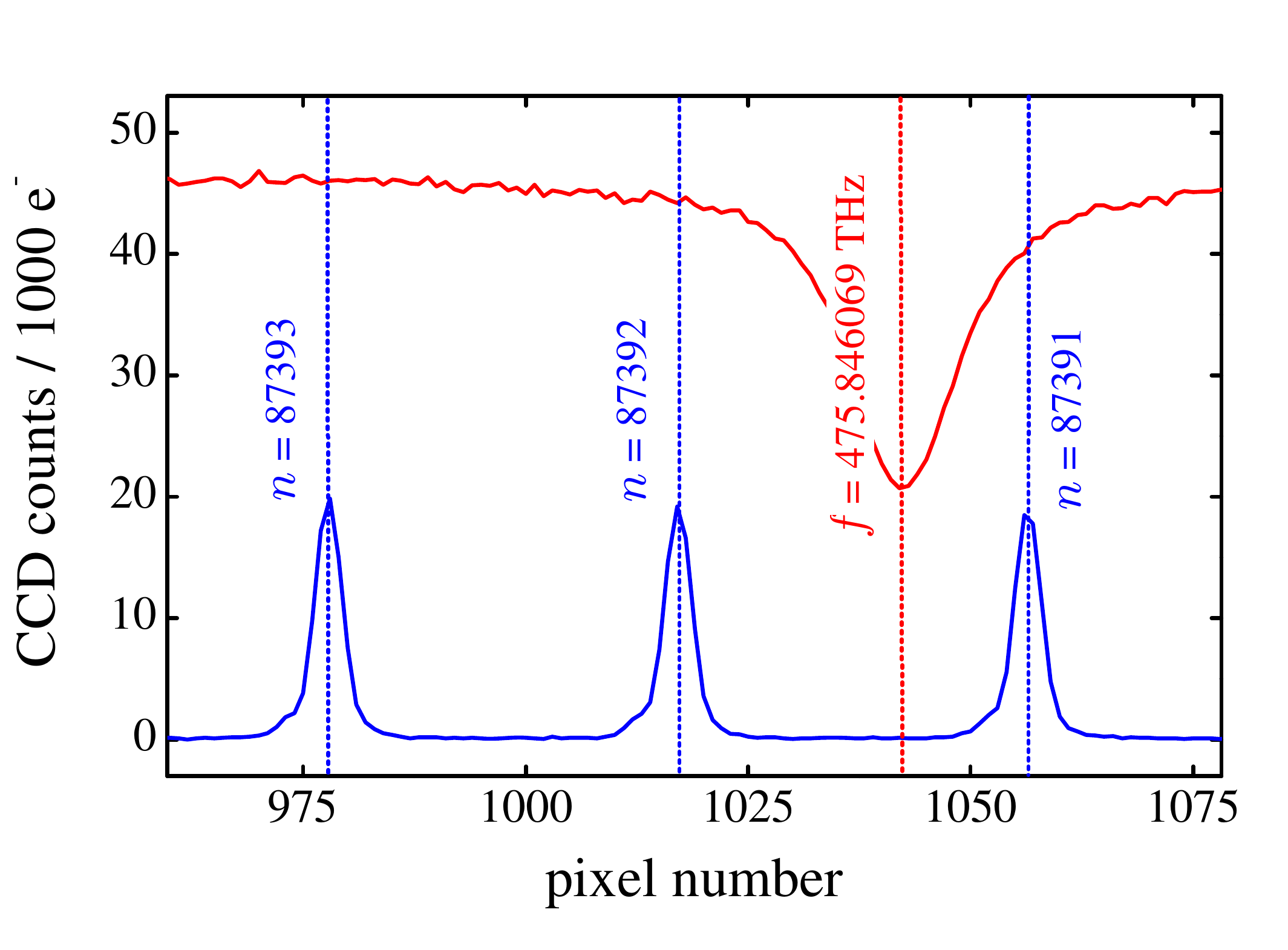}\end{indented}

\caption{\label{fig:Line-identification}Identification of comb lines. The
graph shows a small portion of the comb spectrum (blue solid line
near bottom) and of the solar spectrum (red solid line near top).
The mode numbers $n$ of the comb lines at the positions indicated
by the blue dashed lines are derived from the known frequency $f$
of a telluric $\mathrm{O_{2}}$ line (red dashed line).}
\end{figure}

The known offset frequency and mode spacing of the comb allow to attribute
absolute frequencies to the observed comb lines, as soon as they have
been assigned mode numbers. The latter can be derived from the known
frequencies of the lines in the observed solar spectrum. The telluric
$\mathrm{O_{2}}$ lines are particularly well suited for this task,
because they are fairly deep and much more narrow and stable than
lines originating from the solar photosphere. Their frequencies are
listed in the HITRAN database~\cite{HITRAN}, which we use to attribute
mode numbers to the comb lines as shown in figure~\ref{fig:Line-identification}.
The result is that the 101 observed comb lines have mode numbers ranging
from 87\,417 to 87\,317.

\begin{figure}
\begin{indented}\item[]\includegraphics[width=13.8cm]{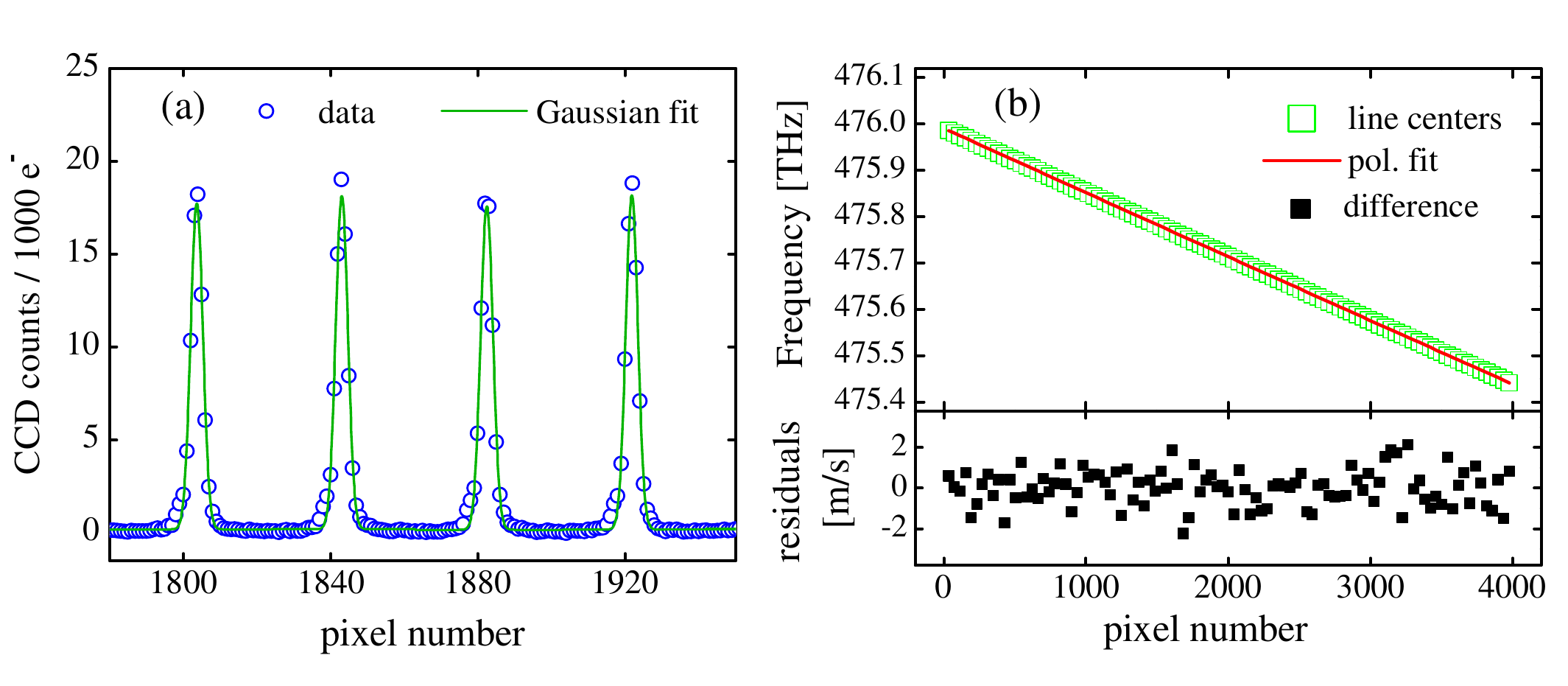}
\end{indented}

\caption{\label{fig:Absolute-calibration}(a) Small section of the observed
comb spectrum, with a Gaussian function fitted to each comb line.
The centers of the fitted Gaussians are assumed as the centers of
the comb lines on the CCD. (b) Upper plot: Frequencies of the observed
comb modes versus their line centers on the CCD. The data points are
approximated by a third-order polynomial. Lower plot: Fit residuals,
converted into m/s. Root-mean-square (RMS) of the fit residuals: 0.87~m/s.
Photon noise: 0.60~m/s.}
\end{figure}

As a next step, the comb lines are fitted with Gaussian functions
using the Levenberg-Marquardt algorithm~\cite{NumRep} (see figure~\ref{fig:Absolute-calibration}(a)).
The fit algorithm delivers uncertainties of the line centers from
error propagation of the photon noise at each data point. The photon
noise is 
assumed to be $\left[\left|N\right|+\left(R\right)^{2}\right]^{1/2}$,
where $N$ denotes the number of detected photoelectrons, and $R$
the CCD readout noise. Subtraction of the dark and bias current of
the CCD is included in $N$, as well as correction for different sensitivity
of adjacent pixels (flat fielding). The centers of the fitted lines
and their uncertainties are the basis for further analysis. The fact
that the observed line profiles are not exactly Gaussian is uncritical
for the further analysis, which merely requires a precise definition
of the line center that leads to reproducible results.

An absolute calibration can be obtained from the line centers on the
CCD and the known frequencies of the corresponding comb modes. With
this information, the pixel-to-frequency relation can be mapped across
the CCD, which is approximated by a third-order polynomial (figure~\ref{fig:Absolute-calibration}(b)).
According to the fitted polynomial, which is hereafter referred to
as the frequency solution, an average CCD pixel spans 138.2~MHz or
87.09~m/s. The RMS deviation of the fit residuals of 0.87~m/s is
close to the photon noise limit of 0.60~m/s, but the discrepancy
still requires an explanation. Using a higher-order polynomial for
fitting does not significantly reduce the discrepancy. In~\cite{Wilken10},
remarkable deviations of the comb line centers from the frequency
solution were observed, and attributed to irregularities of the pixelation
caused by the manufacturing process of the CCD. In particular,
abrupt changes of the pixel-to-frequency relation were found to occur
every 512 pixels, which is equal to the width of the mask used for
fabrication. In our case, the deviations are far less severe and do
not display an obvious, periodically occurring pattern. However, after
averaging the residuals from the frequency solution over all ($>$2700)
comb spectra, their RMS still amounts to 0.50~m/s. Hence, the deviations
above photon noise seem to be largely reproducible and of systematic
nature. We thus attribute the largest portion of these deviations
to irregularities in the CCD pixelation, even though the CCD was apparently
not stitched together from smaller lithographic blocks of pixels,
which is clearly preferable.

\subsection{Calibration repeatability}

To assess the calibration repeatability that the LFC provides in different
calibration schemes, we consider only exposures with pure comb light,
circumventing any limitations from the solar observation.

\begin{figure}
\begin{indented}\item[]\includegraphics[width=12cm]{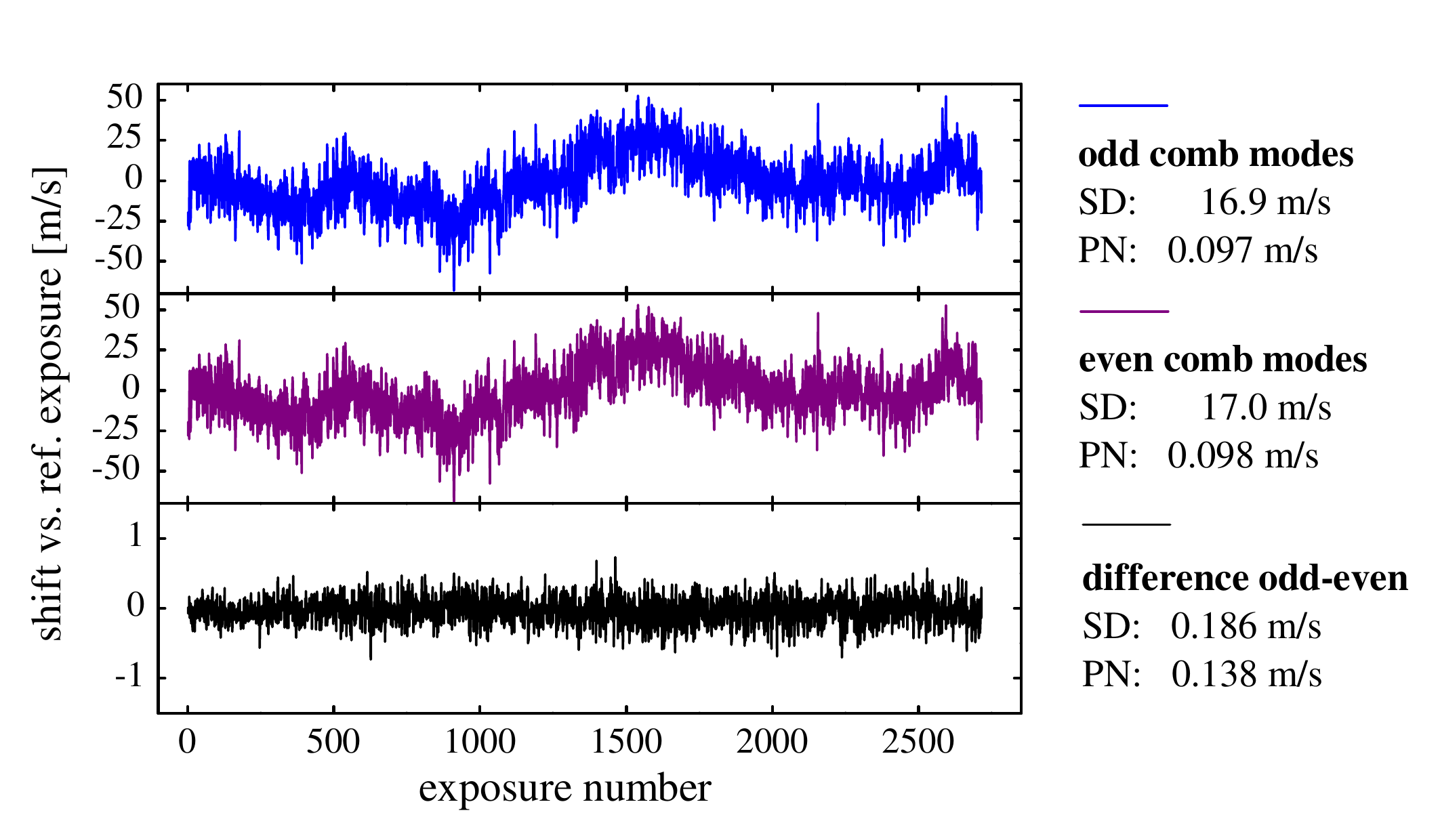}\end{indented}

\caption{\label{fig:odd-even-series}Test of the calibration repeatability
for overlaid calibration. The calibration shifts of two subsets of
modes are compared, one formed by the lines with odd mode numbers
(upper plot), and the other one formed by the lines with even mode
numbers (central plot). The differential calibration shifts between
the two subsets are shown in the lowest plot (notice the different
vertical scale there). Next to each plot, the standard deviation (SD)
and the photon noise limit (PN) of the series are quoted. Integration
time: 1~s. Cadence: 1~frame every 6~s.}
\end{figure}

For testing the calibration repeatability provided by the aforementioned
concept of an overlaid calibration, the observed comb lines are divided
into two groups: One group is represented by the modes with odd mode
numbers, and the other one by even mode numbers. We thus decompose
the LFC into two interleaved combs with twofold mode spacing, and
assess their relative stability. This mimics an overlaid calibration,
with light from two sources simultaneously sent through the same fiber.
For each exposure, we compute the shift of the comb lines relative
to their position in a reference exposure, which is the first exposure
of the series. The combined shift of all odd (even) modes is computed
for every exposure by averaging the individual line shifts. The averaging
of lines applies photon noise weighting, which means that each line
is weighted by its inverse variance given by photon noise. Figure~\ref{fig:odd-even-series}
shows the result for the odd modes, the even modes, and the difference
between the two. While the shifts of the odd modes and the even modes
mainly reflect spectrograph drifts, the difference between the two
should not be larger than photon noise, if the two calibrations are
consistent to that level. The standard deviation of both the odd and
the even modes amounts to 17~m/s, whereas the standard deviation
of the difference is only of 0.186~m/s (photon noise: 0.138~m/s).

\begin{figure}
\begin{indented}\item[]\includegraphics[width=9.5cm]{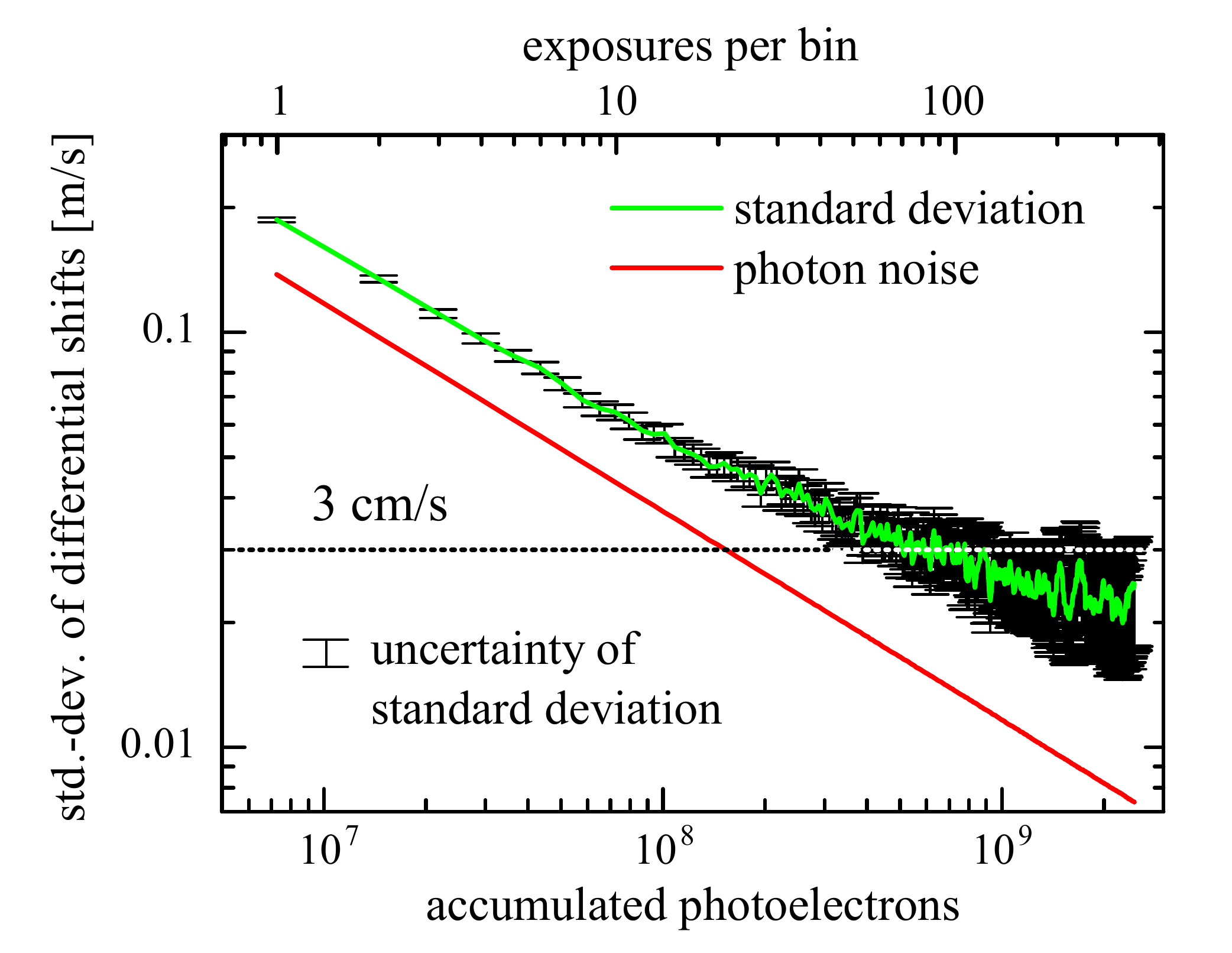}\end{indented}

\caption{\label{fig:Binning-series}Improvement of the calibration repeatability
for overlaid calibration with increasing number of accumulated CCD
counts. The number of counted photoelectrons is increased by summing
up exposures into bins of increasing size (from just 1~exposure per
bin to up to 360). The standard deviation (green line) of the differential
shifts between odd and even comb modes is computed for each series
of binned exposures. The uncertainty of the standard deviation (black
error bars) increases with increasing number of accumulated counts
because of the decreasing number of bins within the series. An exposure
of 1~s was taken once every 6~s.}
\end{figure}

The deviation from photon noise limited statistics of the differential
shifts, unlike in the case of the deviations from the frequency solution,
can hardly be explained by some type of systematic uncertainty. Rather,
we conclude that the scatter above photon noise is due to an excess
noise source. Thermal noise certainly contributes to the excess noise,
since the CCD was not cooled down to the very low temperatures reached
by other astronomical CCDs (typically $-$90 \celsius{} to $-$120 \celsius{}). The largest
portion of the excess noise is however most likely to be provoked
by internal seeing within the spectrograph beam path, due to convection
of air. The convection elements have different air temperature, and
therefore different refractive index, which disturbs the beam. This
occurs because the VTT spectrograph is neither evacuated nor thermally
stabilized. Nevertheless, for this calibration scheme, the improvement
over the passive stability of the spectrograph amounts to about two
orders of magnitude.

Even if the calibration repeatability obtained in a single exposure
is limited by noise, it can still be further enhanced by combining
several exposures into one. If an effect cannot be averaged down in
this manner, it means that its influence displays some systematic
behavior instead of being purely random. We test this by combining
a given number of subsequent exposures into bins. We then carry out
the same analysis as previously for the series of binned exposures.
The result as a function of bin size is shown in figure~\ref{fig:Binning-series}.
The standard deviation drops below 3.0~cm/s for bins of $>$126~exposures.
Again, the photon noise limit is not reached, which would allow an
uncertainty as low as 1.2~cm/s at this point. The standard deviation
seems to form a plateau near 2.3~cm/s. Strikingly, a plateau at approximately
the same value was also found with an LFC on the two-channel spectrograph
HARPS~\cite{Wilken12}. As one of the world's most stable spectrographs,
HARPS is passively stable to better than 1~m/s over at least many
hours~\cite{Mayor03}. The VTT spectrograph in contrast, being neither
evacuated nor temperature or pressure stabilized, can drift by up
to 100~m/s over a day. The results thus show that with science light
and calibration light simultaneously sharing the same spatial mode,
spectrograph stabilization is of reduced importance, provided that
seeing within the spectrograph is kept low enough.

\begin{figure}
\begin{indented}\item[]\includegraphics[width=10cm]{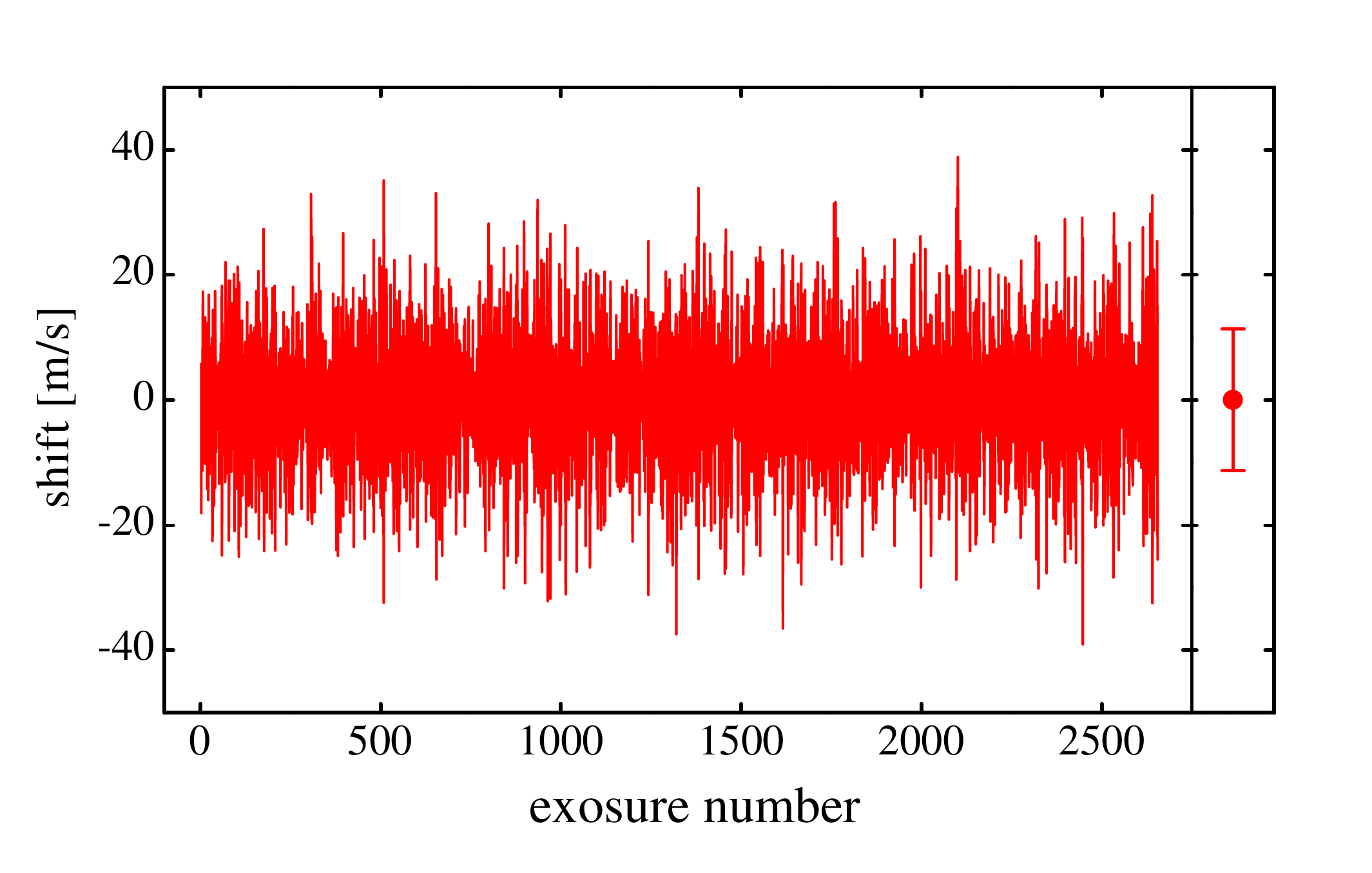}
\end{indented}

\caption{\label{fig:Chopped_comb}Test of the calibration repeatability for
time-interlaced calibration. Each calibration is compared to the interpolation
of the preceding and subsequent one. The photon-noise weighted difference
yields the calibration shifts as shown by the line plot. The error
bar on the right represents the standard deviation of 11.3~m/s of
the measured shifts. The photon noise limit is 0.097~m/s. Integration
time: 1~s.~Cadence:~1~frame~every~6~s.}
\end{figure}

We now turn to the repeatability as determined for time-interlaced
calibration. In this calibration scheme, a science exposure is preceded
and followed by a calibration exposure. From the two calibration exposures,
two frequency solutions are derived, that are interpolated to calibrate
the science exposure. To test this concept with comb light only, each
calibration exposure is compared to the interpolation of the calibrations
6~s before and 6~s after it. The total calibration shift is computed
as the photon-noise weighted average shift of the lines versus the
interpolated calibration. The result is shown in figure~\ref{fig:Chopped_comb}.
It displays a standard deviation of the calibration shifts of 11~m/s,
which is only a rather modest improvement over the 17~m/s as obtained
when considering the fully uncompensated spectrograph drifts (see
upper two plots of figure~\ref{fig:odd-even-series}). The uncertainties
in this calibration scheme are known to arise from spectrograph drifts
that occur on a sub-second time scale~\cite{Doerr12SPIE}. We expect
a slightly better repeatability than demonstrated here for the calibration
of the solar spectra, because of the twofold lower time span between
the interpolated calibration exposures. Generally, the higher the
frame rate, the better is the repeatability that we expect for this
calibration scheme, as shown in~\cite{Doerr12SPIE} using a 10~Hz
frame rate.

\section{Calibrating solar spectra\label{sec:Calibrating-solar-spectra}}

\subsection{Fitting procedures and noise limitations}

\begin{figure}
\begin{indented}\item[]
\includegraphics[width=10cm]{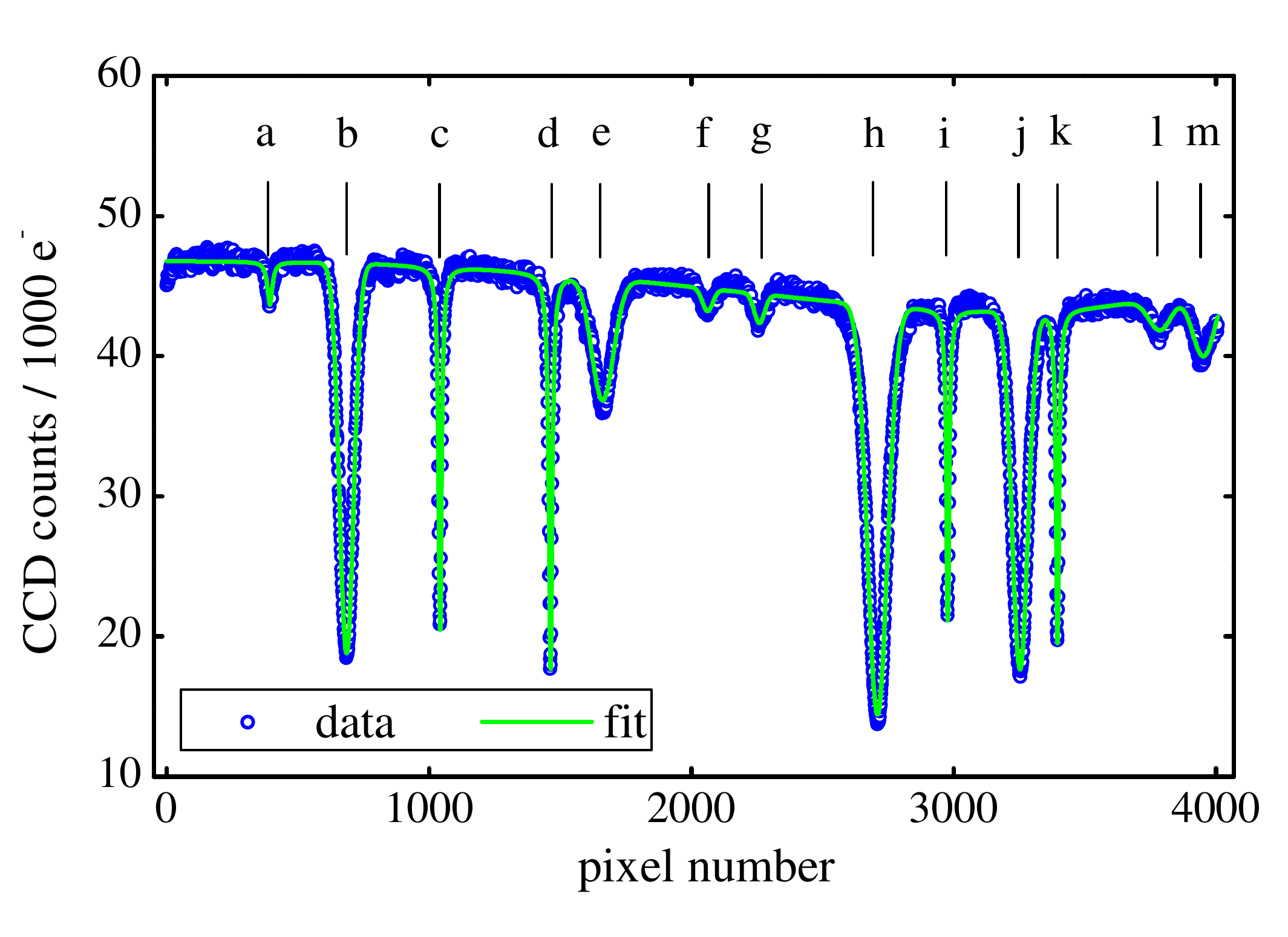}
\end{indented}

\caption{\label{fig:Fit-sun}Fit of the solar spectrum. The solar lines are
fitted with Gaussian functions, while the telluric lines are fitted
with Lorentzian functions. The global descent of the signal level
from left to right is approximated by a fourth-order polynomial. The
black dashes indicate the positions of fitted spectral lines labeled
with the identifiers from table~\ref{tab:PhotonNoise}.}
\end{figure}

We now show how our calibration methods can be applied to solar spectra.
At first, we only consider exposures containing pure sunlight. The
line centers of the absorption lines are determined by fitting, similar
to what was done in section~\ref{sub:Spectrograph-calibration} for
the comb lines. We approximate the solar lines with Gaussian functions,
and the telluric lines with Lorentzian functions. Contrary to the
analysis of the comb exposures, we do not perform the fit on a line-by-line
basis, but fit the sum of the functions of all lines. The overall
signal level is not constant along the echelle order, which is mainly
because of the blaze function of the spectrograph grating. This is
modeled by multiplying the sum of functions by a fourth-order polynomial,
whose coefficients are adjusted by the fit. The result is shown in
figure~\ref{fig:Fit-sun}.

The positions of the line centers obtained in this way can be converted
into frequencies using the frequency solution obtained from time-interlaced
calibration. This permits to track frequency changes (or RV changes)
of the lines. Fitting a line model that accurately reflects the true
line shapes is not needed here, since we are not interested in an
accurate determination of absolute line frequencies. If the absolute
frequency of a solar line needs to be specified, it is customary to
determine its bisector, to account for the asymmetry of the line profile.

Determining line shifts in the overlaid exposures proves to be considerably
more difficult. Here it is a strict necessity that the fit functions
accurately reflect the true line shapes. This is not only the case
for the solar lines, but for all lines, including the comb lines.
If the model used for fitting does not fully match the measured lines,
the fit residuals can influence the fit of the overlaid lines, and
distort their measured line center. Since the relative signal strength
of the two overlaid sources is not fixed, but can vary widely, the
crosstalk between the two channels is not stable. Only if the fit
functions perfectly match the measured line shapes can it be excluded,
that the two fits influence one another in a systematic way. However,
the two overlaid spectra will always be affected by each other's photon
noise: The added photons from the comb bring in additional noise to
the solar spectrum, but do not increase the signal of solar spectrum.
Thus, the overlaid calibration usually comes at the price of decreasing
the signal-to-noise ratio of the spectrum to be calibrated, and also
the calibration itself is affected by a degraded signal-to-noise.

Because the line profiles have to be modeled so accurately, the approach
of fitting simple mathematical functions to the spectral lines is
given up when it comes to analyzing overlaid spectra. Instead, we
use template spectra (as described below) for the Sun and for the
comb, whose sum (including an offset) is used to approximate the overlaid
spectra. The sum of the two templates is matched to the overlaid spectra
by shifting each template's position relative to the CCD pixels and
by scaling its signal level. The square-sum of the fit residuals ($\chi^{2}$),
applying photon noise weighting of the data points, is calculated
by evaluating the templates at the positions of the pixels using cubic
spline interpolation. The fit is designed such that the comb lines
are scaled individually, since their relative signal strength is also
subject to slight changes. Only a selected region around each line
in the solar spectrum is fitted at a time. In this manner, lines in
the solar spectrum can be tracked separately. Especially the telluric
lines are expected to experience shifts that are quite different from
the lines of solar origin. They also change in depth over the course
of the day, owing to the varying air-mass along the line of sight.

\begin{figure}
\begin{indented}\item[]\includegraphics[width=11cm]{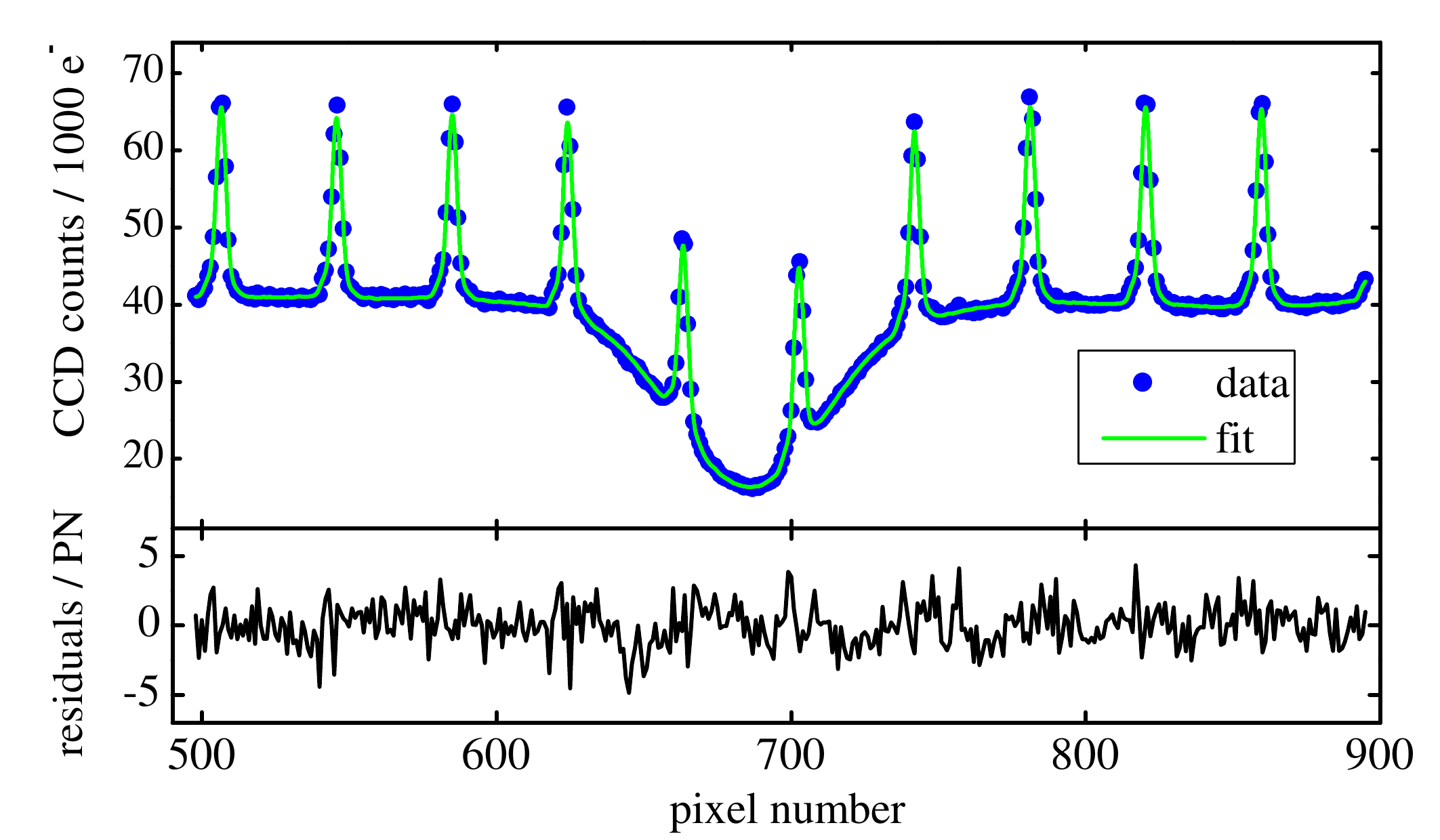}\end{indented}

\caption{\label{fig:Fit-combsun}Fit of a combined Sun and comb spectrum around
a Fe~I solar line. Upper plot: The measured data are fitted with
a sum of templates for the solar spectrum and for the comb spectrum.
Lower plot: Difference between measured data and model fit, expressed
in terms of the photon noise (PN) of the data. The root-mean-square
(RMS) of the distribution should be 1 if the model fits the lines
within the limits given by the photon noise. The actual RMS of the
fit residuals is 1.47 times the photon noise.}
\end{figure}

In a first approach, a single comb exposure and a single solar exposure
were used as templates for fitting. This however resulted in a very
inaccurate localization of the line centers. The reason is, that if
a noisy model is fitted to a noisy data set, the resulting $\chi^{2}$
as a function of the fit parameters acquires finely structured noise,
and thus loses its smoothness. This creates numerous local minima
in the $\chi^{2}$ landscape for the fit algorithm to get stuck in.
The effect was found to be most pronounced for the broad solar lines,
because their shallow slopes of the $\chi^{2}$ valley can more easily
acquire local minima through noise. For the comb, the effect turned
out to be close to negligible, because of its narrow and numerous
lines.

It was thus found to be essential to create templates with minimized
noise. We therefore averaged the Sun and comb templates over all available
comb and solar spectra. Each spectrum was shifted to match the line
centers as indicated by the Gaussian/Lorentzian fits, before including
it into the average. This was done separately for every line in the
solar spectrum. In order to suppress the noise ever further, the templates
for the lines in the solar spectrum were low-pass filtered by a fast-Fourier
transform (FFT) filter. The cut-off of the filter had to be adjusted
carefully depending on the line width, in order to achieve optimal
noise suppression while still fully preserving the line shape. This
yielded extremely smooth line templates especially for broad solar
lines, where the noise suppression is most critical. Due to the narrow
line width of the comb lines, the noise filter was neither applicable
nor needed for the comb.

Figure~\ref{fig:Fit-combsun} shows the fit and its residuals for
a solar Fe~I line. The quality of the fit can be judged from the
distribution of the fit residuals normalized to the photon noise.
The RMS of the residuals is 47~\% above the photon noise of the data
points, which is about equal to the usual level of excess noise, that
was already observed and discussed in section~\ref{sec:Comb-calibration}.
The method therefore fulfills our expectations. All exposures with
combined comb and sunlight are analyzed in this way to track the lines
in the solar spectrum with respect to the overlaid comb.

\begin{table}
\caption{\label{tab:PhotonNoise} Photon noise of the fitted line centers in
the solar spectrum. PN: Photon noise without overlaid comb. PN2: Photon
noise with overlaid comb. The increase of photon noise induced by
the overlay is listed in the rightmost column.}

\medskip{}

\begin{indented}\lineup\item[]%
\begin{tabular}{lllrrr}
\toprule 
identifier & species & origin & PN {[}m/s{]} & PN2 {[}m/s{]} & increase\tabularnewline
\midrule
a & $\mathrm{H_{2}O}$ & telluric & 20.80 & 23.08 & 11 \%\tabularnewline
b & Fe I & solar & \03.17 & \03.68 & 16 \%\tabularnewline
c & $\mathrm{O_{2}}$ & telluric & \01.70 & \01.74 & 2 \%\tabularnewline
d & $\mathrm{O_{2}}$ & telluric & \01.57 & \01.62 & 3 \%\tabularnewline
e & Si I & solar & 14.09 & 15.97 & 13 \%\tabularnewline
f & O I / Ni I & solar & 58.73 & 67.83 & 16 \%\tabularnewline
g & Sc II & solar & 38.92 & 45.33 & 16 \%\tabularnewline
h & Fe I & solar & \03.26 & \03.58 & 10 \%\tabularnewline
i & $\mathrm{O_{2}}$ & telluric & \01.94 & \02.23  & 15 \%\tabularnewline
j & Fe I & solar & \03.51 & \03.86 & 10 \%\tabularnewline
k & $\mathrm{O_{2}}$ & telluric & \01.79 & \01.83 & 2 \%\tabularnewline
l & Fe I & solar & 53.33 & 56.61 & 6 \%\tabularnewline
m & Ti I & solar & 31.98 & 35.92 & 12 \%\tabularnewline
\bottomrule
\end{tabular}\end{indented}
\end{table}

A theoretical performance limitation is the photon noise of the solar
lines. Even though the photon noise can always be reduced by collecting
more exposures, it is useful to estimate the photon noise of the lines
in a single exposure with the two calibration schemes. To this aim,
we employ Monte-Carlo simulation, adding simulated photon noise to
our templates for the solar spectrum to create artificial measurements.
25\,000 simulated exposures were created in this manner, and subsequently
fitted with the templates. The scatter of the line centers over the
series reveals the photon noise of the line positions. For comparison,
an analogous simulation was done for the overlay of comb and Sun.
Table~\ref{tab:PhotonNoise} lists the results. The characteristic
lines all have a photon noise in the low m/s range. The $\mathrm{O_{2}}$
lines have particularly low photon noise, because they are quite deep
and narrow. The overlaid comb increases the photon noise of the solar
lines by 10 to 16~\%. It is striking though, that for the $\mathrm{O_{2}}$
lines the increase can be as low as 2~\%. The reason is, that the
$\mathrm{O_{2}}$ lines are narrow enough to fit between the comb
lines, where they are hardly affected by the comb's photon noise.
The impact of the comb is most pronounced if a comb line is located
on the steepest slope of a line, which is most decisive for the localization
of the line center.

\subsection{Tracking solar lines}

We demonstrate the usefulness of our calibration methods by detecting
signatures of global solar oscillations in our time series of disk-integrated
solar spectra. Most of the power of these oscillations is concentrated
at frequencies of about 3~mHz, which is why they are often referred
to as the "5-minute oscillations". In integrated sunlight, only
low-order oscillations with total velocity amplitudes of the order
of 1~m/s are observed. Sensitive instruments with a stable wavelength
calibration are required to detect these low-amplitude signals. Such
global oscillations at a similar rate and amplitude are present in
the spectra of other stars, and have to be temporally averaged for
the detection of low-mass exoplanets. The measurement of the well-explored
global solar oscillations, whose detection has first been published
in 1979~\cite{Claverie79}, serves as a proof-of-concept for our
technique. Future applications may then investigate more contemporary
problems such as the influence of sunspots on radial velocity detection,
which is of great interest for high-sensitivity exoplanet searches.

For detecting the solar oscillations, we track the average shift of
the three strong Fe~I lines, which are the most distinct features
from the Sun within the spectral range of our observation. Besides
the solar lines, we also track the shifts of the $\mathrm{O_{2}}$
lines imprinted by Earth's atmosphere. These lines are often used
as a low-precision reference, but are restricted to certain spectral
regions. Their line centers are affected by Doppler shifts through
changing wind speeds and directions. At noon, they are expected to
be stable within a few m/s. This is because when the Sun is near the
zenith, the winds are approximately perpendicular to the line of sight.
During the course of the afternoon, however, the stability of the
lines is supposed to degrade, as the component of the wind speeds
along the line of sight increases.

\begin{figure}
\begin{indented}\item[]\includegraphics[width=13.8cm]{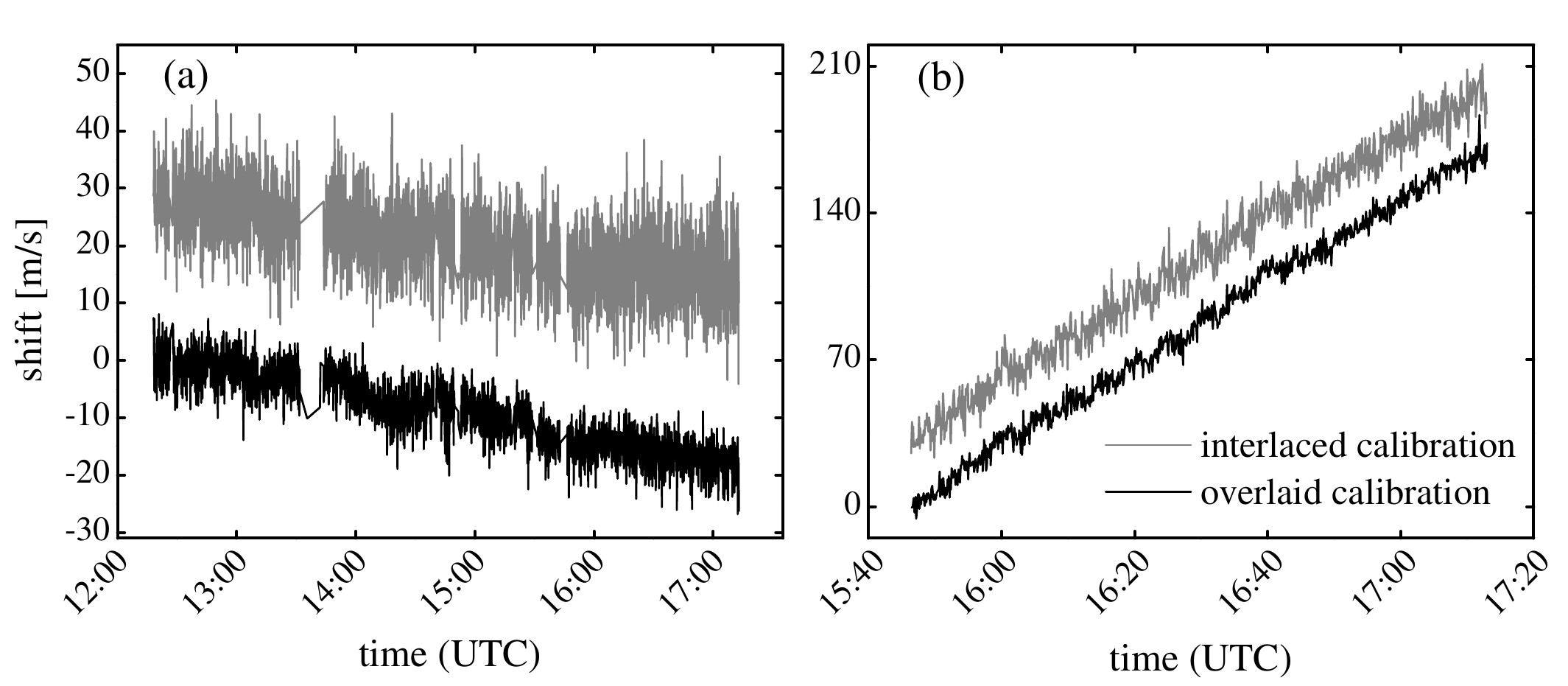}\end{indented}

\caption{\label{fig:Tracking-lines}Tracking lines in the solar spectrum using
time-interlaced calibration (gray lines) and overlaid calibration
(black lines). For clarity, the gray lines were given a vertical offset
relative to the black lines. (a)~Shift of the telluric lines with
time, averaged over four $\mathrm{O_{2}}$ lines. (b)~Average shift
of the three strongest Fe~I lines.}
\end{figure}

Time traces of the average shifts of the four $\mathrm{O_{2}}$ lines,
and of the three Fe~I lines, are shown in figure~\ref{fig:Tracking-lines},
both with time-interlaced and overlaid calibration. The time series
is longer for the $\mathrm{O_{2}}$ lines than for the Fe~I lines,
since for the tracking of the $\mathrm{O_{2}}$ lines, telescope guiding
problems can be tolerated. Over the course of the series, all four
$\mathrm{O_{2}}$ lines experience about the same drift of approximately
$-$17~m/s, similar to what found in earlier investigations~\cite{Balthasar82,Figueira10}.
The largest portion of the drift is assumed to be caused by winds~\cite{Balthasar82,Figueira10},
with a weaker contribution arsing from the changing air-mass along
the line of sight~\cite{Caccin85,Figueira10}. Time-interlaced calibration
confirms the results obtained with overlaid calibration, but with
2$-$3~times larger scatter. In the case of the overlaid spectra, the
scatter is dominated by the measurement of the solar and telluric
lines, rather than by the calibration. This limitation could be reduced
by averaging a larger number of lines, e.g. with a spectrograph that
records a wider spectral range.

\begin{figure}
\begin{raggedleft}
\includegraphics[width=14cm]{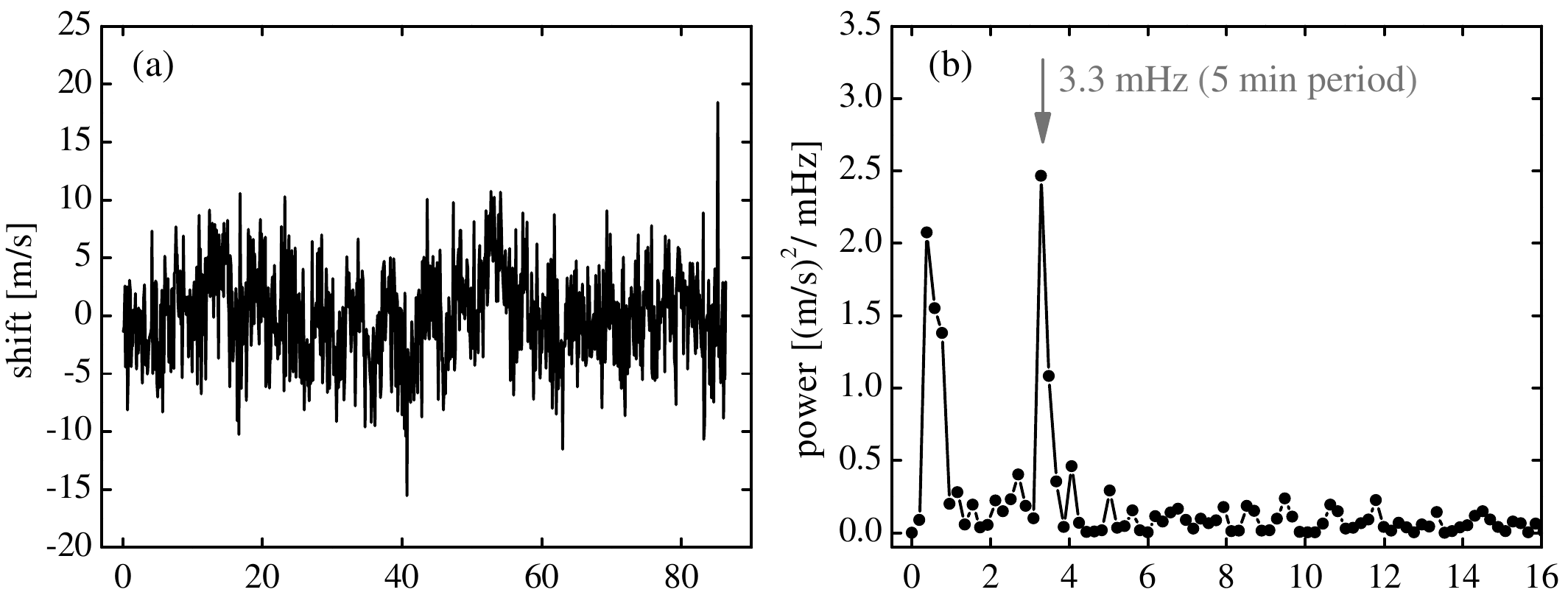}
\par\end{raggedleft}

\begin{raggedleft}
\includegraphics[width=14cm]{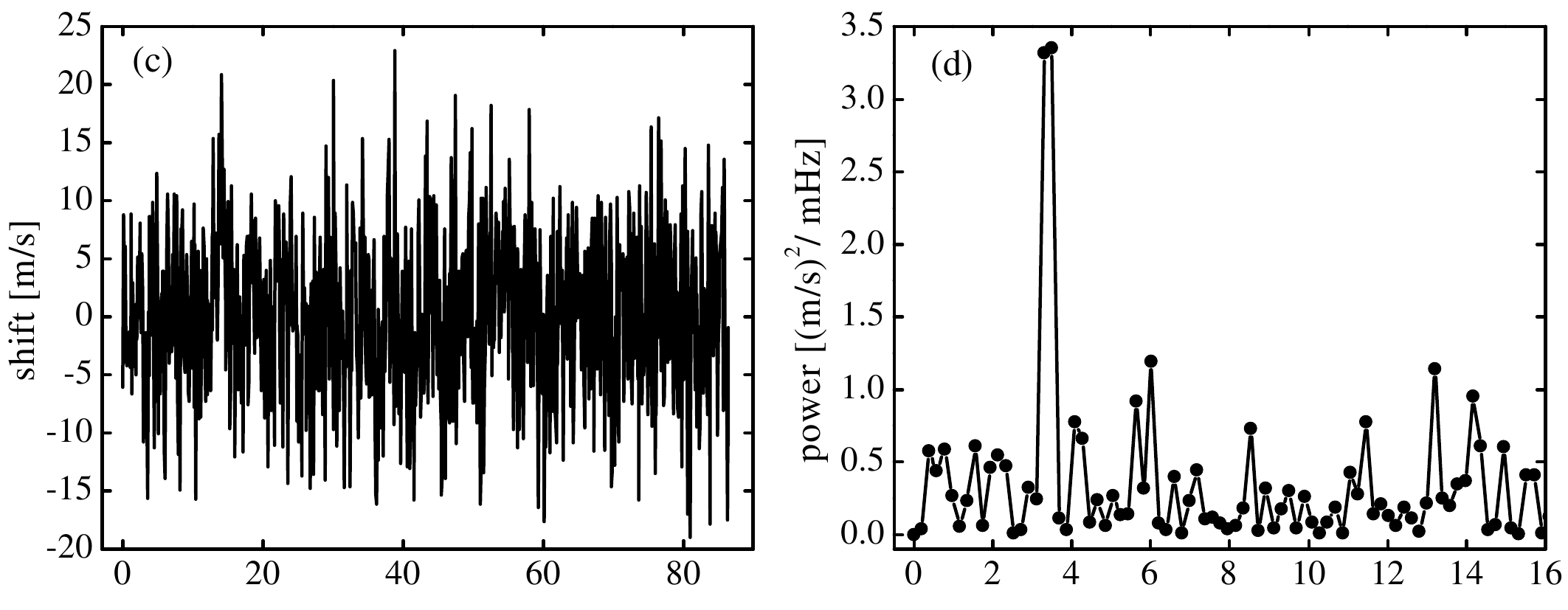}
\par\end{raggedleft}

\begin{raggedleft}
\includegraphics[width=14cm]{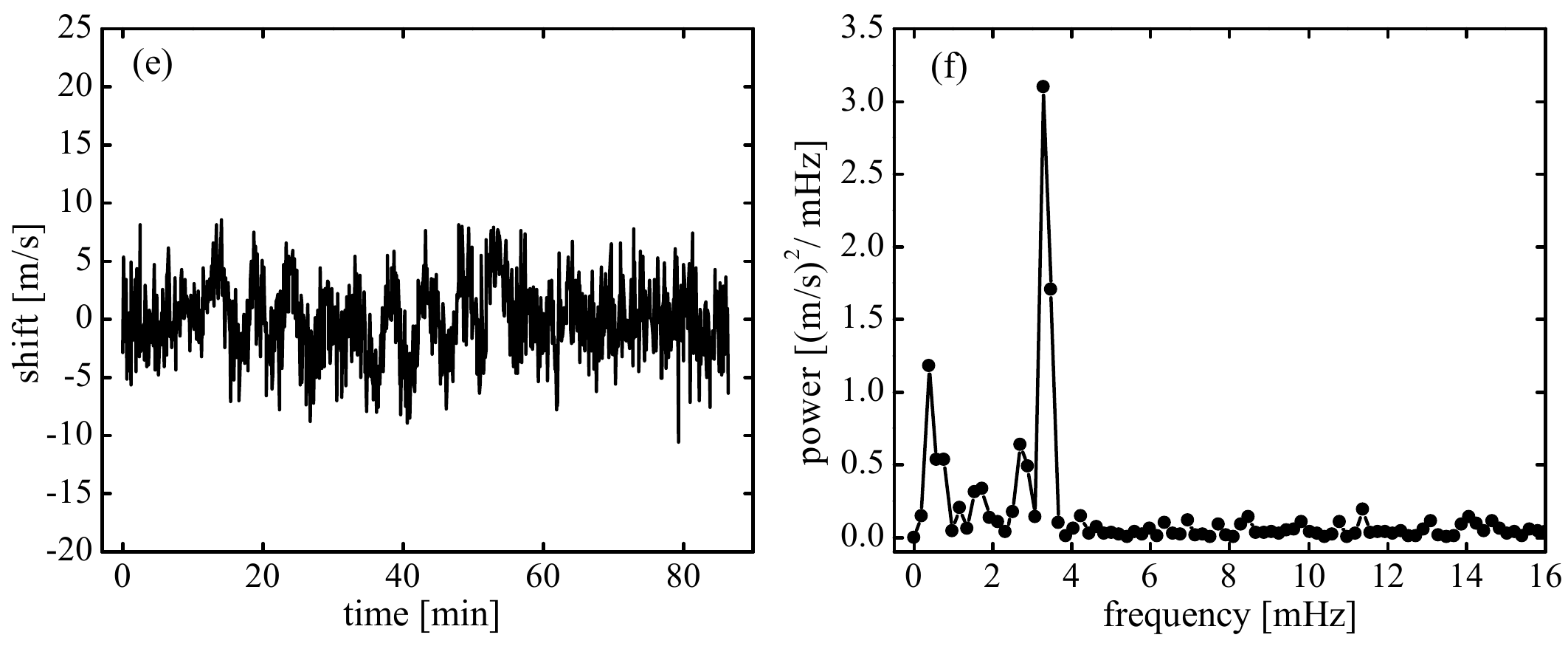}
\par\end{raggedleft}

\caption{\label{fig:Detection-of-p-modes}Detection of solar oscillations using
the three strongest Fe~I lines of figure~\ref{fig:Fit-sun} / table~\ref{tab:PhotonNoise}.
Graphs on the left: Time traces with a second-order polynomial subtracted.
Graphs on the right: Power spectra computed by fast Fourier transformation
of the time traces. Note that these spectra scale quadratically in
amplitude. Top: Overlaid calibration. Center: Time-interlaced calibration.
Bottom: Calibration with telluric $\mathrm{O_{2}}$ lines.}
\end{figure}

The time traces of the Fe~I lines, as shown in figure~\ref{fig:Tracking-lines},
display a strong drift from the rotation of the Earth, which needs
to be subtracted. We do this by fitting and subtracting a second-order
polynomial to the time traces, leading to the curves shown in figure~\ref{fig:Detection-of-p-modes}~(a)
and~(c). This "quick-and-dirty" way of
accounting for Earth's rotation is sufficient for the aim of detecting
solar oscillations, and partly also compensates other unwanted effects
such as differential extinction of the solar disc~\cite{Davies14}
and telescope guiding errors. Besides the expected 5-minute oscillations,
the time traces also display a slower component, which is probably
caused by guiding errors. The graphs on the right-hand side of figure~\ref{fig:Detection-of-p-modes}
represent the power spectra of the time traces to their left, computed
by FFT. The 5-minute oscillations are represented by a peak at 3.3~mHz
in the power spectra. The frequency and the signal strength is in
good agreement with the results reported in~\cite{Claverie79}. Our
series of acquisitions is not long enough in time to fully resolve
the substructure of the peak at 3.3~mHz. However, the weak side-lobes
next to the central peak are likely to stem from partly resolved oscillation
modes. The total oscillation power of the central peak at 3.3~mHz
corresponds to an amplitude of 88~cm/s. Amplitudes of the globally
averaged solar oscillations have also been characterized in~\cite{Kjeldsen08}
using the HARPS spectrograph, were the strongest component is reported
to have an amplitude of 25~cm/s, confirming earlier findings. Since
in our measurements, the peak at 3.3~mHz contains numerous unresolved
components, our results are in unison with this. Supposed errors from
telescope guiding appear below 1~mHz. It is not clearly seen in the
power spectrum obtained from time-interlaced calibration, probably
because it interferes with noise. 

The peak detection threshold with time-interlaced and overlaid calibration
can be estimated by regarding the noise floor of the power spectrum.
For an oscillation signal of unresolved width and a 3~sigma confidence
level, it is of 37~cm/s for overlaid calibration. The calibration
tests in section~\ref{sec:Comb-calibration} yielded somewhat lower
values with this scheme, but there only the standard deviation was
specified, corresponding to merely 1~sigma. The guiding errors deteriorate
the sensitivity of the measurement only at frequencies of below 1~mHz,
where a signal would be masked by the guiding errors if its amplitude
is below 63~cm/s. For time-interlaced calibration, the 3~sigma peak-detection
threshold amounts to 69~cm/s. The performance is better than in section~\ref{sec:Comb-calibration}
because of the higher frame rate, and because considering the time
series through a frequency filter is a way of applying a temporal
average.

On the time scale of the oscillations, the $\mathrm{O_{2}}$ lines
are relatively stable, and can thus also be used as a reference for
their detection. Figure~\ref{fig:Detection-of-p-modes}~(e) and
(f) shows the analysis of the time trace of the Fe~I lines as referenced
to the $\mathrm{O_{2}}$ lines. All findings derived from comb calibration
are confirmed. Due to the lower photon noise, the 3~sigma peak detection
threshold is of 28~cm/s. Our above analysis however clearly reveals
that over longer time frames, the $\mathrm{O_{2}}$ lines are unstable
even on the 10~m/s scale, and are thus unsuited for sensitive detection
of phenomena that occur on longer time horizons.

\begin{figure}
\begin{indented}\item[]
\includegraphics[width=13.75cm]{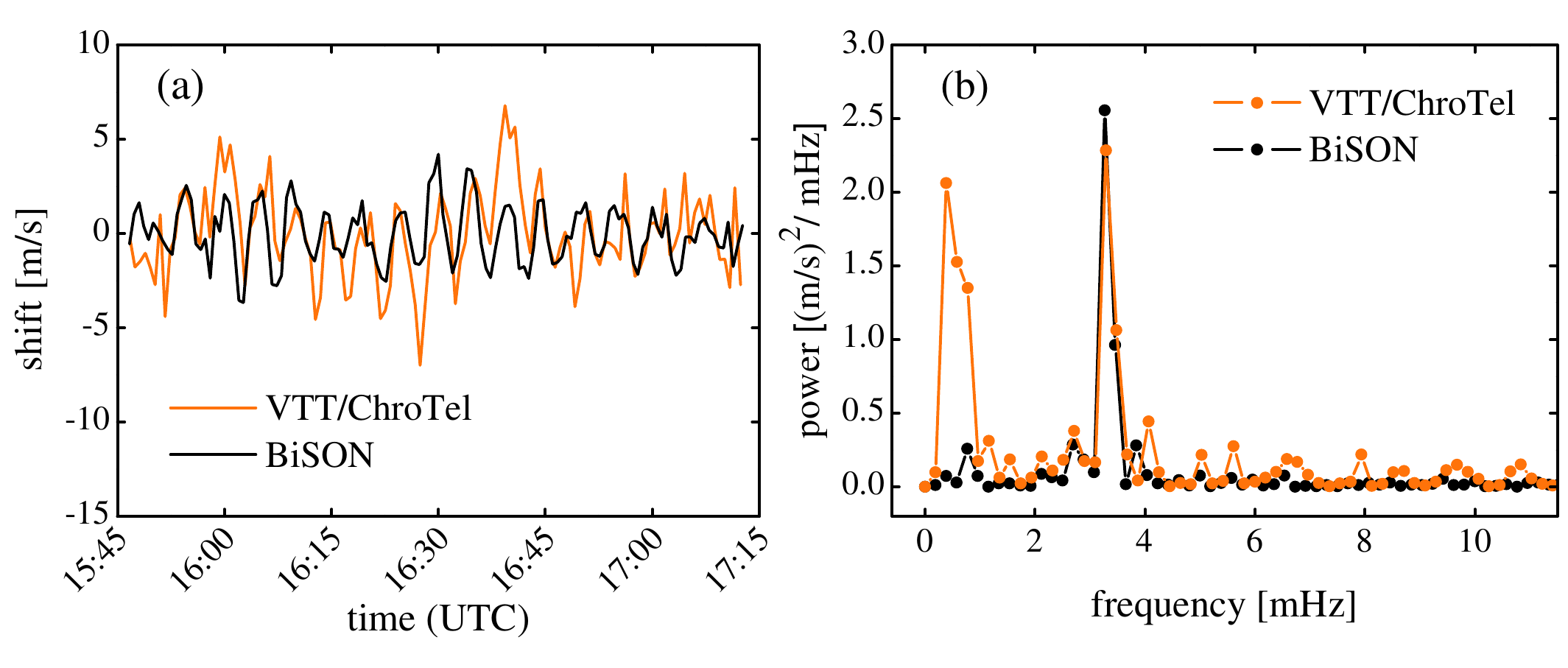}
\end{indented}

\caption{\label{fig:VTT-vs-BiSON}Comparison of solar oscillation measurements
from VTT/ChroTel to BiSON data of the same date and time. The VTT/ChroTel
data were calibrated with the overlaid comb. Additionally, the time
series was binned over 7~exposures (42~s), to match the temporal
resolution of the BiSON data (40~s). (a)~Time series of RV shifts.
(b)~Power spectra, computed by fast Fourier transformation.}
\end{figure}

Disk-integrated solar oscillations have been measured for more than
3~decades by BiSON~\cite{Davies14}. BiSON is a network of telescopes
that observe solar oscillations via resonant scattering of a potassium
line. This is a method very different from ours, which makes a comparison
between the two all the more interesting. BiSON publicly provides
data that also cover the time period of our observation. To these
data, we compare our observation using overlaid calibration with the
LFC. We bin our data over 7~exposures (42~s), to match the temporal
resolution of the BiSON data (40~s). The comparison is shown in figure~\ref{fig:VTT-vs-BiSON}.
The measured phases of the oscillations nicely match each other. Also
the overall powers of the 5-minute oscillations agree within the expected
uncertainties. The most obvious differences come from the guiding
errors in our data, that appear at frequencies below 1~mHz. More
subtle differences might arise from the different weighting of the
solar limb in our observation.

\section{Conclusion}

We have demonstrated two LFC-assisted spectrograph calibration schemes,
that both make use of a multiplexed SMF: Time-interlaced calibration,
where the calibration and the measurement are separated in time, and
overlaid calibration, where comb light is superimposed with light
from the telescope. The latter technique has proven to be superior
due to its simultaneous spatial mode matching between comb and sunlight.
This approach was found to be practical and very useful, but came
at the expense of a more complex data analysis and of a slightly increased
noise level. We could demonstrate a calibration repeatability down
to better than 3.0~cm/s, paralleling the so far best results from
the LFC on the HARPS spectrograph of 2.5~cm/s~\cite{Wilken12}.
The use of SMFs brings in the additional benefit of eliminating modal
noise as present in multimode fibers. With the proof-of-principle
demonstration in this work, consisting of solar oscillation detection
in disk-integrated sunlight, and of tracking of telluric lines, the
concept is now ready for exploring physical phenomena on the Sun.

In astronomy, frequency combs are still a novelty and non-standard
equipment at observatories. This however, is about to change, and
LFC-assisted spectroscopy is envisioned to have a flourishing future
in astronomy. Our present work shows how future astronomical LFCs
could be utilized. Other observatories might consider a multiplexed
fiber delivery, although employing multimode fibers. With further
progress in adaptive optics, or by making use of astrophotonic devices
such as photonic lanterns, future instruments might use multiplexed
SMFs on stars other than the Sun, e.g. for exoplanet searches.

\ack{}{}

We gratefully acknowledge the help of Frank Grupp from the University
Observatory Munich, who has supported us with helpful discussions
and advice. The KIS contribution to this project was in part funded
by the Leibniz-Gemeinschaft within the "Pakt f\"ur Forschung und Innovation".
The VTT is operated by the KIS at the Observatorio del Teide of the
Instituto de Astrof\'isica de Canarias in Tenerife, Spain. LW is supported
by the Forschungsprojekt GZ~788 of Chinesisch-Deutsches Zentrum f\"{u}r
Wissenschaftsf\"{o}rderung, and the Young Researcher Grant of National
Astronomical Observatories, Chinese Academy of Sciences. We thank
the BiSON team (funded by UK Science, Technology and Facilities Council,
STFC) for their friendly permission to use the time series of BiSON
RV measurements. We also thank the technical staff of the VTT for
their assistance.

\section*{References}{}

\bibliographystyle{iopart-num}
\bibliography{bibliography}

\end{document}